\documentclass[lettersize,journal]{IEEEtran}

\usepackage{cite}
\usepackage[T1]{fontenc}
\usepackage[utf8]{inputenc}
\usepackage{booktabs}
\usepackage{algorithm}
\usepackage{float}
\usepackage{titlesec}
\titlespacing{\subsection}{0pt}{4pt}{2pt}
\usepackage{algpseudocode}
\usepackage[most]{tcolorbox}
\usepackage{array}
\usepackage{colortbl}
\usepackage{xcolor}
\usepackage{pifont}

\usepackage{amsmath,amssymb,amsfonts}
\usepackage{titlesec}
\usepackage{tikz}
\usetikzlibrary{arrows.meta, positioning, fit, calc}
\titlespacing{\section}{0pt}{2pt}{2pt}
\titlespacing{\subsection}{0pt}{2pt}{2pt}
\titlespacing{\subsubsection}{0pt}{2pt}{2pt}
\usepackage[switch]{lineno}
\usepackage{booktabs}
\usepackage{tcolorbox}
\usepackage{graphicx}
\usepackage{caption}
\usepackage{graphicx}
\usepackage{textcomp}
\usepackage[table]{xcolor}
\usepackage{colortbl}
\usepackage{algorithm}
\usepackage{algpseudocode}
\usepackage{tabularx}
\usepackage{listings}
\usepackage{mdframed}
\usepackage{tikz}
\usetikzlibrary{positioning,arrows.meta,fit,backgrounds}
\definecolor{softblue}{RGB}{232,240,254}
\definecolor{softviolet}{RGB}{240,232,250}
\definecolor{softorange}{RGB}{252,236,218}
\definecolor{softgreen}{RGB}{230,244,234}
\definecolor{lineblue}{RGB}{91,133,217}
\definecolor{lineviolet}{RGB}{141,110,179}
\definecolor{lineorange}{RGB}{214,145,70}
\definecolor{linegreen}{RGB}{92,160,98}
\usepackage{pifont}
\usepackage{tikz}
\usepackage{pgfplots}
\pgfplotsset{compat=1.18}

\usetikzlibrary{arrows.meta, positioning}

\usetikzlibrary{arrows.meta,positioning,fit,calc,shapes.geometric,matrix}

\definecolor{headergray}{RGB}{230,230,230}
\definecolor{hyperblue}{RGB}{220,235,255}
\definecolor{good}{RGB}{220,245,220}
\definecolor{bad}{RGB}{255,220,220}

\definecolor{successgreen}{RGB}{0,120,0}
\definecolor{failred}{RGB}{180,0,0}
\definecolor{codegray}{RGB}{240,240,240}

\lstdefinestyle{svsuccess}{
  language=Verilog,
  backgroundcolor=\color{codegray},
  basicstyle=\ttfamily\footnotesize,
  keywordstyle=\color{successgreen}\bfseries,
  commentstyle=\color{successgreen},
  numbers=none,
  frame=single,
  breaklines=true
}

\lstdefinestyle{svfailure}{
  language=Verilog,
  backgroundcolor=\color{codegray},
  basicstyle=\ttfamily\footnotesize,
  keywordstyle=\color{failred}\bfseries,
  commentstyle=\color{failred},
  numbers=none,
  frame=single,
  breaklines=true
}

\definecolor{LogBlue}{RGB}{25,70,160}
\definecolor{LogGreen}{RGB}{0,120,60}
\definecolor{LogRed}{RGB}{180,30,30}
\definecolor{LogGray}{RGB}{90,90,90}
\definecolor{LogPurple}{RGB}{120,60,160}
\definecolor{LogBG}{RGB}{248,248,248}

\lstdefinestyle{hhlog}{
  basicstyle=\ttfamily\scriptsize,
  columns=fullflexible,
  keepspaces=true,
  breaklines=true,
  breakatwhitespace=false,
  frame=single,
  framerule=0.3pt,
  rulecolor=\color{black},
  showstringspaces=false,
  xleftmargin=0.6em,
  xrightmargin=0.6em,
  aboveskip=4pt,
  belowskip=2pt,
  backgroundcolor=\color{LogBG},
  linewidth=\columnwidth,
  keywordstyle=\bfseries\color{LogBlue},
  commentstyle=\color{LogGray},
  alsoletter={=,-,.,_,>,<,/,\\,?,:,*,+},
  morekeywords={PHASE,P1,P2,Correctness,SA,DC,iter,T,pick,score,compile,sim,warn,ACCEPT,REJECT,best_score,saved,area,power,wns,SELECTED}
}

\lstdefinestyle{compilerlog}{
  basicstyle=\ttfamily\footnotesize,
  columns=fullflexible,
  keepspaces=true,
  breaklines=true,
  frame=single,
  rulecolor=\color{black},
  showstringspaces=false,
  aboveskip=6pt,
  belowskip=6pt
}

\def\BibTeX{{\rm B\kern-.05em{\sc i\kern-.025em b}\kern-.08em
    T\kern-.1667em\lower.7ex\hbox{E}\kern-.125emX}}

\begin{document}

\title{MoSAIC: Learning DAG Scheduling Heuristics from Optimal Schedules through Motif-Aware Recovery and LLM-Guided Reasoning}

\author{
Shiva Ahir\textsuperscript{1},
Hashmath Shaik\textsuperscript{1},
Alex Doboli\textsuperscript{1},
Nadeem Yaseen\textsuperscript{2},
Simona Doboli\textsuperscript{3}\\[6pt]
\small \textsuperscript{1}Department of Electrical \& Computer Engineering,
Stony Brook University, New York, United States of America\\
\small \textsuperscript{2}Barcelona Supercomputing Center,
Barcelona, Catalonia, Spain\\
\small \textsuperscript{3}Department of Computer Science,
Hofstra University, New York, United States of America\\[4pt]
\small \texttt{\{shiva.ahir,hashmath.shaik,alex.doboli\}@stonybrook.edu}\\
\small \texttt{nadeem.yaseen@bsc.es},
\texttt{simona.doboli@hofstra.edu}
}

\maketitle
\pagestyle{empty}
\thispagestyle{empty}
% For double-blind review, keep author block empty
%\author{}

\maketitle

\begin{abstract}
Efficient task scheduling in heterogeneous computing systems remains a fundamental challenge in embedded and high-level synthesis flows. This paper presents MoSAIC (Motif-aware Scheduling with AI-driven Cognition), a learning-based framework that bridges the gap between exact and heuristic scheduling approaches for DAG-based applications. The system uses the reasoning capabilities of large language models (LLMs) to learn the relationships between DAG structures and scheduling priority functions. The framework first leverages the OR-Tools CP-SAT solver \cite{ref04} to generate optimal schedules, which are then used to supervise the learning of compact priority-based scheduling functions through Bayesian Optimization. These learned heuristics capture relationships between scheduling decisions and structural graph features such as rank, depth, and fanout, enabling the construction of a reusable knowledge base across diverse DAG instances. Experimental results across synthetic graph families and real-world benchmarks, including PolyBench and MachSuite, show that the learned heuristics consistently improve over classical approaches such as HEFT \cite{ref02}, while maintaining low runtime overhead. The results indicate that combining optimal supervision with learning-based generalization provides a practical approach for scalable and adaptive task scheduling.
\end{abstract}


\section{Introduction}

Efficient task scheduling for Directed Acyclic Graphs (DAGs) is a fundamental problem in heterogeneous computing systems, particularly in embedded systems, high-level synthesis (HLS), and parallel processing workflows \cite{ref02, ref23, ref25}. The goal is to assign tasks to processors and determine their execution order such that overall execution time (makespan) is minimized while satisfying precedence and communication constraints. This problem is NP-hard, motivating the development of heuristic and approximate solutions~\cite{ref01}, with comprehensive treatments available in scheduling literature \cite{ref15}.

To achieve optimality, optimization-based approaches such as Integer Linear Programming (ILP) have been explored~\cite{ref03}. However, ILP formulations scale poorly and quickly become computationally infeasible for larger DAGs. Recent advances in constraint programming, particularly the CP-SAT solver~\cite{ref04}, and industrial solvers like IBM ILOG CP Optimizer \cite{ref15}, \cite{ref16} have showcased strong performance in solving scheduling problems optimally for small-to-medium-sized instances, offering a practical alternative to traditional ILP approaches.

Classical scheduling algorithms such as HEFT and CPOP~\cite{ref02}, along with force-directed scheduling (FDS)~\cite{ref24} and other heuristics evaluated in comparative studies~\cite{ref14}, remain widely used due to their efficiency and low computational complexity. These approaches rely on handcrafted priority functions derived from structural properties of the DAG, such as upward rank, computation cost, and communication overhead. While effective in practice, these heuristics are static and often fail to generalize across diverse graph topologies and system configurations. New applications and domains are likely to introduce DAGs with novel structural patterns, which necessitates the design of new priority functions. This process, however, is cumbersome and requires manual effort.

In parallel, learning-based approaches have gained attention for improving scheduling decisions using data-driven techniques~\cite{ref05, ref06, ref07}. These methods aim to learn scheduling policies or heuristics that adapt to varying workloads and system characteristics. However, many existing approaches either require extensive training or lack interpretability which makes it difficult to extract meaningful scheduling strategies. Prior work on network motifs has shown that recurring structural patterns can capture essential characteristics of complex graphs~\cite{ref10}. This insight motivates the use of structural patterns to guide scheduling decisions, an opportunity underused in traditional algorithms.

In this work, we propose MoSAIC (Motif-aware Scheduling with AI-driven Cognition), a learning-driven framework that bridges optimal scheduling and heuristic design. MoSAIC first utilizes the CP-SAT solver \cite{ref04} to generate optimal schedules for a diverse set of DAGs. These optimal schedules are then used as supervision to recover priority-based scheduling functions via Bayesian Optimization, capturing the relationship between graph structural features such as rank, depth, fanout, and communication patterns and effective scheduling strategies. This process enables the construction of a structured dataset mapping DAG characteristics to high-quality priority functions. Building on this dataset, MoSAIC further explores the use of Machine Learning (ML) models, including Large Language Models (LLMs), to predict scheduling heuristics for unseen graphs. At inference time, the predicted priority function can be used within a standard list scheduling framework to produce near-optimal schedules with minimal computational overhead.

We evaluate MoSAIC on both synthetic DAGs~\cite{ref11} (Erdős--Rényi, Barabási--Albert, Watts--Strogatz) and real-world benchmarks (PolyBench~\cite{ref08} and MachSuite~\cite{ref09}). Experimental results suggest that MoSAIC consistently outperforms classical heuristics such as HEFT, achieving makespan close to optimal while maintaining low runtime overhead.

A key observation underlying this work is that effective scheduling behavior is not invariant across task graphs. Different structural patterns induce distinct priority preferences that cannot be captured by a single fixed heuristic. Instead, scheduling can be viewed as a structured mapping from graph topology to decision policies. This perspective indicates that optimal scheduling behavior can be learned, represented, and transferred across graphs by conditioning on structural features and motifs. MoSAIC is built on this principle, combining optimal supervision, heuristic recovery, and learning-based generalization into a unified framework. The structure of the paper is as follows: Section~II presents related work, Section~III introduces the problem formulation, and Section~IV presents the MoSAIC framework. Section~V presents the experimental setup and evaluation, followed by a discussion of observations, limitations, and the conclusion.

\section{Related Work}

Task scheduling for Directed Acyclic Graphs (DAGs) has been studied for many years, especially in systems where different processors work together. Most existing methods fall into three main categories: heuristic-based methods, optimization-based methods, and learning-based methods.

Heuristic methods like HEFT (Heterogeneous Earliest Finish Time) and CPOP (Critical Path on a Processor)~\cite{ref02} assign priorities to tasks based on properties such as task importance, computation cost, and communication delays. These methods are fast and widely used because they are simple to implement. However, they follow fixed rules, which means they do not adapt well when the structure of the graph changes.

To find the best possible schedules, some works model the problem using optimization techniques like Integer Linear Programming (ILP)~\cite{ref03}. These methods can produce optimal results by considering all constraints carefully. More recently, constraint programming approaches such as the CP-SAT solver~\cite{ref04} have improved performance and can solve moderately sized problems efficiently. Even so, these methods are still expensive to run and are not practical for large graphs or real-time use.

Learning-based approaches have been explored to make scheduling more adaptive. Techniques such as reinforcement learning and neural combinatorial optimization~\cite{ref18, ref19} allow systems to learn scheduling strategies from data instead of relying on fixed rules~\cite{ref05, ref06}. Research in machine learning for combinatorial optimization~\cite{ref07}, \cite{ref22} also shows that models can learn to handle complex decision-making problems. However, these approaches often require a lot of training data, take time to train, and can be hard to interpret.

In addition to this, research on network motifs shows that repeating patterns in graphs can reveal important structural information~\cite{ref10}. These patterns have been useful in understanding complex networks, but they have not been widely used for designing scheduling strategies.

The existing work either focuses on fast heuristics, exact optimization, or learning-based approaches. However, combining these ideas, especially using optimal solutions to guide the design of reusable and interpretable scheduling strategies—has not been fully explored.

\section{Problem Formulation}

Task scheduling in heterogeneous computing systems is represented using a directed acyclic graph (DAG) 
$G=(V,E)$, where each node $v \in V$ denotes a task and each directed edge $(u,v)\in E$ represents a precedence dependency. A task cannot begin execution until all of its parent tasks have completed and any required data has been communicated. The target platform consists of a heterogeneous processor set $P=\{p_1,p_2,\dots,p_m\}$, where the execution time of a task depends on the processor to which it is assigned.

A schedule is defined by two components: a processor assignment $\pi(v)$ for each task and a start time $s(v)$. Let $w(v,\pi(v))$ denote the execution cost of task $v$ on processor $\pi(v)$, and let $c(u,v)$ denote the communication cost incurred when dependent tasks are placed on different processors. A feasible schedule must satisfy precedence and resource constraints. The precedence relation may be written as:
\begin{equation}
\begin{aligned}
s(v) \;\ge\;& s(u) + w(u,\pi(u)) \\
&+\, c(u,v)\,\mathbf{1}\!\left[\pi(u) \neq \pi(v)\right], \\
&\forall\ (u,v) \in E
\end{aligned}
\label{eq:precedence}
\end{equation}
where $\mathbb{I}[\cdot]$ is the indicator function. In addition, no two tasks assigned to the same processor may overlap in time.

The objective is to minimize the completion time of the final task, that is, the makespan:
\begin{equation}
C_{\max} = \max_{v\in V}\left(s(v)+w(v,\pi(v))\right)
\end{equation}
This optimization problem is NP-hard because task ordering, processor assignment, and communication-aware execution interact combinatorially ~\cite{ref01}.

Classical approaches address this problem in two very different ways. Exact methods search for the globally optimal solution \cite{ref17}, but their computational cost grows rapidly with graph size. Heuristic methods, such as list scheduling, are efficient, but they depend heavily on fixed priority rules that may not remain effective across different graph structures, as observed in comparative studies~\cite{ref14}. This creates a gap between optimality and scalability. A method is therefore needed that preserves the structural quality of exact schedules while retaining the execution speed of heuristics.

The central idea is to represent scheduling behavior through a learnable priority function rather than a fixed handcrafted rule. Task priority is computed using the parametric function defined in Section~III. To obtain meaningful parameter values, high-quality reference schedules are first used as supervision. For each benchmark DAG, an optimal schedule, defined by processor assignment $\pi^*$ and start times $s^*$, is computed offline:
\begin{equation}
(\pi^*, s^*) = \arg\min_{\pi,s} C_{\max}
\end{equation}
These schedules are not used directly at deployment time; instead, they provide supervision about what effective scheduling behavior looks like under heterogeneous constraints. The role of learning is to recover a compact priority rule whose induced list schedule closely approximates the optimal schedule, thereby preserving solution quality while maintaining low runtime overhead.

This recovery process is written as:
\begin{equation}
\theta^*=\arg\min_{\theta}\left|C_{\max}^{H(\theta)}-C_{\max}^{*}\right|
\end{equation}
where $C_{\max}^{H(\theta)}$ is the makespan obtained by list scheduling under the priority function $H(\cdot;\theta)$, and $C_{\max}^{*}$ is the optimal makespan. Bayesian Optimization is used to search the parameter space efficiently \cite{ref20}, since the objective is non-convex and expensive to evaluate repeatedly.

Once such priority functions are recovered across many graphs, scheduling is no longer viewed only as an isolated combinatorial optimization problem. It becomes a structured prediction problem in which graph characteristics are used to infer an effective scheduling rule. In this way, exact optimization provides supervisory signal, while priority-based scheduling provides an interpretable and scalable execution mechanism. The result is a formulation that connects optimal scheduling quality with fast heuristic inference through learned structural adaptation.

\section{MoSAIC Framework Methodology}

MoSAIC operationalizes the above formulation through a structured multi-stage pipeline consisting of three stages: (i) constructing supervision from exact schedules, (ii) distilling this supervision into compact heuristic rules, and (iii) leveraging the resulting data to enable learning-based generalization. The CP-SAT supervision stage is illustrated in Fig. 2. The framework is designed so that expensive optimization is confined to an offline stage, while inference on unseen DAGs remains lightweight.

The first stage is benchmark construction. A structurally diverse corpus of DAGs is required so that the learned scheduling behavior is not tied to a single dependency pattern. For this reason, the framework includes both synthetic and real application-derived task graphs. Synthetic DAGs provide controlled variation in topology and allow systematic exposure to different structural regimes. Graph families such as Erd\H{o}s--R\'enyi, Barab\'asi--Albert, and Watts--Strogatz \cite{ref11} capture different connectivity characteristics, including random dependency spread, hub-dominated organization, and clustered locality. Real benchmark DAGs  \cite{ref08, ref09} complement this by introducing dependency structures that arise from practical computation kernels.

The second stage is structural characterization. Each task $v \in V$ is associated with a feature vector:
{\small
\begin{equation}
\mathbf{f}(v)=
\left[
\begin{array}{l}
w(v),\ \mathrm{rank}_u(v),\ \mathrm{depth}(v),\\
\mathrm{fanout}(v),\ \mathrm{indegree}(v),\ \bar{c}_{\mathrm{out}}(v)
\end{array}
\right]
\label{eq:feature_vector}
\end{equation}
}
where these features capture both computational cost and structural position in the DAG. Motifs are extracted from each DAG using the gSpan frequent subgraph mining 
algorithm \cite{ref25} applied to 5-node weakly connected induced subgraphs with at least 
4 directed edges. Several structural patterns are observed in task graphs. Among these, three recurring structural motif classes are shown in Fig.~1. These motifs capture common patterns of task divergence and convergence that inform the design of effective scheduling policies.

\begin{figure}[t]
\centering
\begin{tcolorbox}[
    colback=gray!5,
    colframe=black,
    boxrule=0.4pt,
    arc=6pt,
    width=\columnwidth,
    left=2pt,right=2pt,top=4pt,bottom=4pt
]
\centering
\resizebox{\columnwidth}{!}{%
\begin{tikzpicture}[
    every node/.style={font=\scriptsize},
    motifnode/.style={circle, draw=black, thick, minimum size=6mm, inner sep=0pt},
    forkjoin/.style={motifnode, fill=violet!25},
    fanout/.style={motifnode, fill=green!25},
    fanin/.style={motifnode, fill=orange!30},
    titlebox/.style={font=\bfseries\scriptsize},
    >=Latex,
    thick
]

% ---------- Fork-Join ----------
\node[titlebox] at (0,1.8) {Fork-Join};

\node[forkjoin] (f1) at (0,0.9) {1};
\node[forkjoin] (f2) at (-0.8,0) {2};
\node[forkjoin] (f3) at (0,0) {3};
\node[forkjoin] (f4) at (0.8,0) {4};
\node[forkjoin] (f0) at (0,-0.9) {0};

\draw[->] (f1) -- (f2);
\draw[->] (f1) -- (f3);
\draw[->] (f1) -- (f4);
\draw[->] (f2) -- (f0);
\draw[->] (f3) -- (f0);
\draw[->] (f4) -- (f0);

% ---------- Fan-Out ----------
\node[titlebox] at (3.8,1.8) {Fan-Out};

\node[fanout] (o4) at (3.8,0.9) {4};
\node[fanout] (o0) at (2.6,0) {0};
\node[fanout] (o1) at (3.4,0) {1};
\node[fanout] (o2) at (4.2,0) {2};
\node[fanout] (o3) at (5.0,0) {3};

\draw[->] (o4) -- (o0);
\draw[->] (o4) -- (o1);
\draw[->] (o4) -- (o2);
\draw[->] (o4) -- (o3);

% ---------- Fan-In ----------
\node[titlebox] at (7.7,1.8) {Fan-In};

\node[fanin] (i1) at (6.7,0.9) {1};
\node[fanin] (i2) at (7.4,0.9) {2};
\node[fanin] (i3) at (8.1,0.9) {3};
\node[fanin] (i4) at (7.4,0) {4};
\node[fanin] (i0) at (7.4,-0.9) {0};

\draw[->] (i1) -- (i4);
\draw[->] (i2) -- (i4);
\draw[->] (i3) -- (i4);
\draw[->] (i4) -- (i0);

\end{tikzpicture}
}
\end{tcolorbox}

\caption{Representative recurring structural motifs in MoSAIC: fork-join, fan-out, and fan-in.}
\label{fig:motif_classes}
\end{figure}
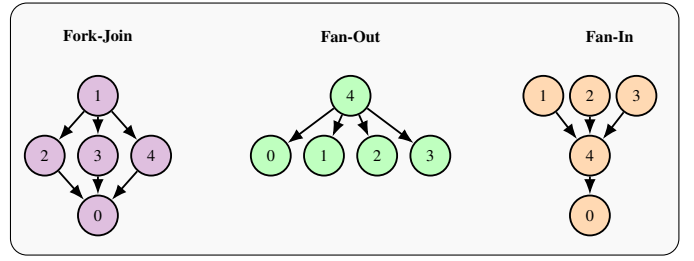

For each DAG, motif detection 
produces a count vector $m(G) \in \mathbb{Z}^{K}$, where $K$ is the number of 
distinct motif classes found across the corpus, and each entry records the 
number of occurrences of that motif class in the graph. This count vector is 
concatenated with the graph-level feature vector $\Phi(G)$ and provided as 
input to both the Bayesian Optimization stage and the LLM fine-tuning stage.

The third stage is exact supervision generation. Each DAG is converted into a constraint-based scheduling formulation and solved using CP-SAT \cite{ref04}. The solution defines a total ordering of tasks along with processor assignments. This induces an optimal decision sequence:
\begin{equation}
\pi^* = (v_1, v_2, \dots, v_{|V|})
\end{equation}
where each decision corresponds to selecting the next schedulable task. This sequence reflects optimal scheduling behavior under precedence and resource constraints and serves as supervision for subsequent stages.

The fourth stage is heuristic recovery. Instead of storing full schedules, MoSAIC computes a compact priority function that approximates the optimal behavior. For a given parameter vector $\theta$, list scheduling produces a makespan $C_{\max}^{H(\theta)}$. The recovery objective is defined as:
\begin{equation}
\mathcal{L}(\theta) = \frac{C_{\max}^{H(\theta)} - C_{\max}^*}{C_{\max}^*}
\end{equation}
which represents the normalized optimality gap. Bayesian Optimization is used to minimize $\mathcal{L}(\theta)$ and obtain $\theta^*$ for each DAG. This step converts exact supervision into a reusable scheduling rule.

The fifth stage is learning-based generalization based on an LLM. The dataset constructed across DAGs consists of tuples:
\begin{equation}
\mathcal{D} = \{ (G_i,\ \Phi(G_i),\ \theta_i^*) \}_{i=1}^N
\end{equation}
where $\Phi(G_i)$ represents graph-level and motif-level descriptors. A predictor is then trained to approximate the mapping:
\begin{equation}
\hat{\theta} = F\big(\Phi(G)\big)
\end{equation}
allowing inference of scheduling policies without solving the underlying optimization problem.

The sixth stage integrates LLM-guided reasoning. Given structural descriptors, motif context, and examples from $\mathcal{D}$, the LLM is prompted to generate scheduling priority preferences. To further align the model with optimal scheduling behavior, reinforcement learning is used. Let a predicted decision sequence be $\hat{\pi}$. A reward signal is defined as:
\begin{equation}
R(\hat{\pi}) = -\frac{C_{\max}(\hat{\pi}) - C_{\max}^*}{C_{\max}^*}
\end{equation}
so that schedules closer to optimal receive higher reward. This feedback guides the model to refine its decision-making 
policy based on schedule quality rather than textual similarity. 
At inference time, the LLM operates as a step-by-step task 
selector: given the serialized ready queue, motif context, and 
current processor state, it directly outputs the next task 
identifier, bypassing the need to explicitly predict $\hat{\theta}$.

\textbf{Example:} Consider a DAG containing a fork-join region followed by a long critical chain. A scheduler that prioritizes only immediate readiness may delay the critical path, increasing makespan. In contrast, structure-aware policies assign higher priority to tasks that influence downstream synchronization. The combination of motif descriptors, recovered priority functions, and reward-driven learning allows the system to capture such behavior.

%The overall flow of the exact supervision stage is shown below.

\begin{figure}[t]
\centering
\resizebox{\columnwidth}{!}{%
\begin{tikzpicture}[
node distance=7mm and 6mm,
box/.style={
draw,
rounded corners=4pt,
align=center,
minimum width=2.55cm,
minimum height=0.88cm,
font=\scriptsize,
inner sep=3pt,
line width=0.9pt
},
arrow/.style={-{Latex[length=2.2mm]}, line width=1.1pt, draw=black!75}
]

% Top row
\node[box, fill=softblue, draw=lineblue] (dag) {Benchmark DAG\\Node weights + Edge weights};
\node[box, fill=softblue, draw=lineblue, right=of dag] (parse) {Parse DAG\\Tasks, dependencies,\\communication costs};
\node[box, fill=softviolet, draw=lineviolet, right=of parse] (model) {CP-SAT Model\\Start/end vars\\Assignment vars};

% Middle row
\node[box, fill=softorange, draw=lineorange, below=of model] (constraints) {Constraints\\Precedence\\NoOverlap\\Makespan objective};
\node[box, fill=softorange, draw=lineorange, left=of constraints] (solve) {CP-SAT Solver\\Branching + propagation\\Optimal search};
\node[box, fill=softgreen, draw=linegreen, left=of solve] (output) {Optimal Schedule\\Task, processor,\\start, end};

% Bottom box
\node[box, fill=softgreen, draw=linegreen, below=of solve, minimum width=2.95cm] (supervision) {Supervision for MoSAIC\\Optimal labels for\\BO and LLM stages};

% Arrows
\draw[arrow] (dag) -- (parse);
\draw[arrow] (parse) -- (model);
\draw[arrow] (model) -- (constraints);
\draw[arrow] (constraints) -- (solve);
\draw[arrow] (solve) -- (output);
\draw[arrow] (solve) -- (supervision);

% Stage labels
\node[font=\scriptsize\bfseries, text=lineblue, above=3pt of parse] {Input Representation};
\node[font=\scriptsize\bfseries, text=lineviolet, above=3pt of model] {Optimization Model};
\node[font=\scriptsize\bfseries, text=lineorange] at ($(constraints.south)!0.6!(solve.south)+(6pt,-12pt)$) {Search \& Constraints};
\node[font=\scriptsize\bfseries, text=linegreen, below=3pt of supervision] {Output Supervision};

\end{tikzpicture}%
}
\caption{CP-SAT stage in the MoSAIC pipeline. A weighted DAG is transformed into a constraint optimization model, solved for an optimal schedule, and used to generate supervision for downstream learning stages.}
\label{fig:cpsat_pipeline}
\end{figure}
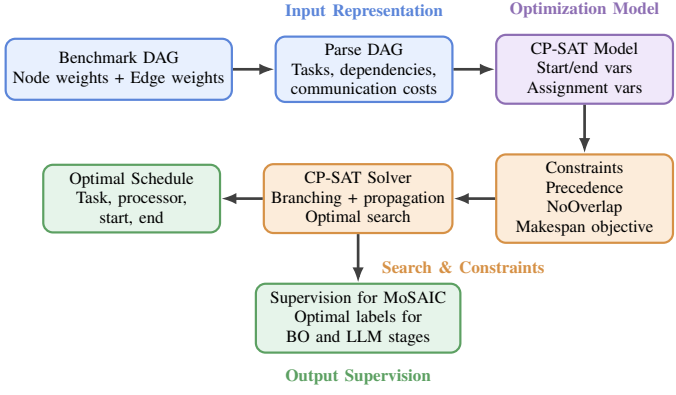
In this organization, each stage has a distinct role. The problem formulation defines what must be optimized and why fixed heuristics are insufficient. The MoSAIC methodology connects benchmark generation, motif analysis, exact supervision, heuristic recovery, and learning-based reasoning into a unified pipeline for near-optimal scheduling on unseen DAGs, as illustrated in Fig.~4. Although the evaluation is conducted on dual-processor settings, the learned priority function is not restricted by processor count and operates on structural task features.

\section{Experimental Setup and Evaluation}

The experimental methodology studied how optimal scheduling behavior could be transferred into efficient and adaptive scheduling policies using learning-based techniques. Instead of repeatedly solving the scheduling problem using expensive optimization methods, the approach treated optimal schedules as a source of supervision from which reusable scheduling strategies could be learned.

The evaluation followed a three-stage pipeline. A key aspect of this evaluation was that it did not treat scheduling performance as an isolated outcome, but as a consequence of structural decision quality. Rather than only comparing final makespan values, the study analyzed how scheduling behavior evolved across different graph families, structural patterns, and decision regimes. This included evaluating generalization across unseen graph distributions by measuring decision-level accuracy of learned policies, and isolating the contribution of each component through ablation. This perspective ensured that improvements were not tied to specific instances, but reflected consistent adaptation to underlying graph structure.

First, a diverse set of DAGs was constructed using both synthetic generation and real benchmark extraction. Second, each graph was solved using the CP-SAT solver to obtain an optimal or near-optimal schedule. Third, these optimal schedules were used to derive interpretable scheduling policies and train learning-based models that could operate on unseen graphs.

%The scheduling problem is modeled using a directed acyclic graph $G = (V, E)$, where each node represents a task and each edge represents a precedence constraint with associated communication cost. A schedule assigns each task $v \in V$ to a processor $\pi(v)$ and determines its start time $s(v)$ while respecting precedence and resource constraints. The objective is to minimize the makespan:
%\begin{equation}
%C_{\max} = \max_{v \in V} \left( s(v) + w(v, \pi(v)) \right)
%\end{equation}

All experiments were conducted on a dual-processor system to maintain a controlled setting while preserving non-trivial scheduling complexity. This configuration is intentionally chosen to enable reliable generation of optimal supervision using CP-SAT, whose computational complexity increases significantly with processor count. By restricting to two processors, the framework ensures consistent and high-quality optimal schedules for training while isolating structural scheduling behavior from hardware-induced variability. Importantly, the learned priority function is independent of processor count and can be directly extended to multi-processor settings using standard list scheduling.

% \begin{figure}[t]
% \centering

% % Insert your existing image or TikZ code here.

% % \vspace{1mm}
% % \caption{Motif-conditioned interpretation of learned priority behavior across graph families.
% % Different local structures induce different scheduling preferences.
% % BA-like hub-dominated regions emphasize critical-path urgency while discouraging premature fanout expansion.
% % ER-like regions balance urgency and topological progress.
% % WS-like clustered and merge-heavy regions assign stronger importance to depth and indegree to reduce synchronization delay.
% % This illustrates that effective scheduling behavior is graph-structure dependent rather than governed by a single fixed priority rule.}
% % \label{fig:motif_priority_interpretation}
% % \end{figure}

\subsection{Datasets}
The benchmark set included both synthetic and real DAGs to capture different aspects of scheduling behavior.

Synthetic DAGs were generated from three graph families: Erd\H{o}s--R\'enyi (ER), Barab\'asi--Albert (BA), and Watts--Strogatz (WS). ER graphs exhibited uniform randomness in connectivity, BA graphs produced hub-dominated structures affecting scheduling due to dependency imbalance, and WS graphs exhibited local clustering with mixed parallelism and synchronization patterns. Training graphs were kept below 150 nodes so that CP-SAT could generate optimal supervision within practical time limits; the learned policy was then deployed on larger graphs where exact solving was infeasible.

Real benchmark DAGs were extracted from application kernels including GEMM, SYRK, FFT, AES, ATAX, BICG, CONV2D, and STENCIL \cite{ref08, ref09}, obtained through a compilation flow that converted program code into dependency graphs capturing computation and communication behavior. These benchmarks reflected realistic execution patterns closer to practical applications.

\subsection{Optimal Supervision and Heuristic Recovery}

For each DAG, the CP-SAT solver was used to compute an optimal schedule. The solver incorporated precedence constraints, processor assignment, and resource constraints to determine task ordering and execution timing that minimized makespan. These schedules were treated as ground-truth supervision.

Instead of using the optimal schedules directly at runtime, the objective was to extract the underlying scheduling behavior in a compact form. For this purpose, each task was represented using structural features such as execution weight, upward rank, depth, fanout, indegree, and communication statistics. These features captured different dimensions of scheduling relevance, including critical-path urgency, dependency structure, and communication impact.

\vspace{0.5em}

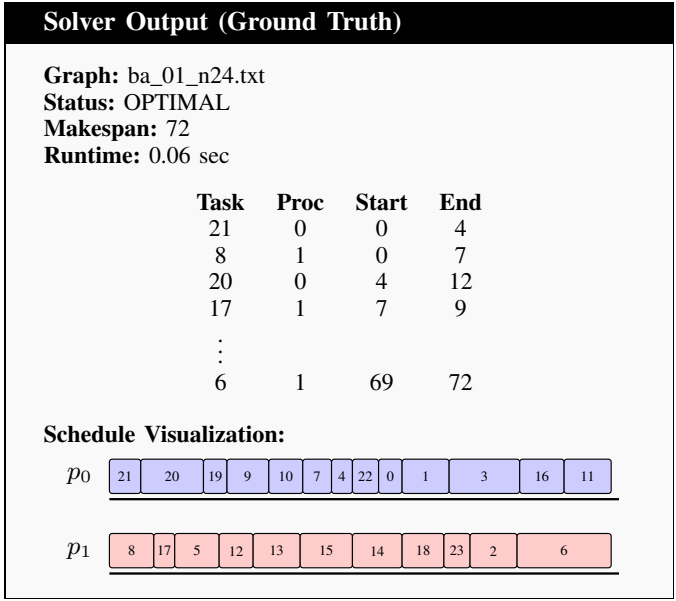
\begin{figure}[t]
\centering

\begin{tcolorbox}[
colback=gray!5,
colframe=black,
title=Solver Output (Ground Truth),
boxrule=0.5pt,
sharp corners,
fonttitle=\bfseries
]

\small
\textbf{Graph:} ba\_01\_n24.txt \\
\textbf{Status:} OPTIMAL \\
\textbf{Makespan:} 72 \\
\textbf{Runtime:} 0.06 sec

\vspace{0.5em}

\begin{center}
\begin{tabular}{cccc}
\textbf{Task} & \textbf{Proc} & \textbf{Start} & \textbf{End} \\
21 & 0 & 0 & 4 \\
8  & 1 & 0 & 7 \\
20 & 0 & 4 & 12 \\
17 & 1 & 7 & 9 \\
\vdots &   &   &   \\
6  & 1 & 69 & 72 \\
\end{tabular}
\end{center}

\vspace{0.5em}
\textbf{Schedule Visualization:}

\begin{center}
\resizebox{0.95\columnwidth}{!}{%
\begin{tikzpicture}[x=0.32cm,y=0.7cm,transform shape]

% Processor labels
\node[anchor=east] at (0.2,2.5) {\small $p_0$};
\node[anchor=east] at (0.2,1.2) {\small $p_1$};

% Baselines
\draw[thick] (0.5,2.1) -- (20.0,2.1);
\draw[thick] (0.5,0.8) -- (20.0,0.8);

% p0 tasks
\draw[fill=blue!20,rounded corners=1pt] (0.5,2.2) rectangle (1.7,2.8) node[midway] {\tiny 21};
\draw[fill=blue!20,rounded corners=1pt] (1.7,2.2) rectangle (4.1,2.8) node[midway] {\tiny 20};
\draw[fill=blue!20,rounded corners=1pt] (4.1,2.2) rectangle (5.0,2.8) node[midway] {\tiny 19};
\draw[fill=blue!20,rounded corners=1pt] (5.0,2.2) rectangle (6.6,2.8) node[midway] {\tiny 9};
\draw[fill=blue!20,rounded corners=1pt] (6.6,2.2) rectangle (7.9,2.8) node[midway] {\tiny 10};
\draw[fill=blue!20,rounded corners=1pt] (7.9,2.2) rectangle (9.0,2.8) node[midway] {\tiny 7};
\draw[fill=blue!20,rounded corners=1pt] (9.0,2.2) rectangle (9.8,2.8) node[midway] {\tiny 4};
\draw[fill=blue!20,rounded corners=1pt] (9.8,2.2) rectangle (10.8,2.8) node[midway] {\tiny 22};
\draw[fill=blue!20,rounded corners=1pt] (10.8,2.2) rectangle (11.7,2.8) node[midway] {\tiny 0};
\draw[fill=blue!20,rounded corners=1pt] (11.7,2.2) rectangle (13.5,2.8) node[midway] {\tiny 1};
\draw[fill=blue!20,rounded corners=1pt] (13.5,2.2) rectangle (16.2,2.8) node[midway] {\tiny 3};
\draw[fill=blue!20,rounded corners=1pt] (16.2,2.2) rectangle (17.9,2.8) node[midway] {\tiny 16};
\draw[fill=blue!20,rounded corners=1pt] (17.9,2.2) rectangle (19.7,2.8) node[midway] {\tiny 11};

% p1 tasks
\draw[fill=red!20,rounded corners=1pt] (0.5,0.9) rectangle (2.2,1.5) node[midway] {\tiny 8};
\draw[fill=red!20,rounded corners=1pt] (2.2,0.9) rectangle (3.0,1.5) node[midway] {\tiny 17};
\draw[fill=red!20,rounded corners=1pt] (3.0,0.9) rectangle (4.7,1.5) node[midway] {\tiny 5};
\draw[fill=red!20,rounded corners=1pt] (4.7,0.9) rectangle (6.0,1.5) node[midway] {\tiny 12};
\draw[fill=red!20,rounded corners=1pt] (6.0,0.9) rectangle (7.8,1.5) node[midway] {\tiny 13};
\draw[fill=red!20,rounded corners=1pt] (7.8,0.9) rectangle (9.8,1.5) node[midway] {\tiny 15};
\draw[fill=red!20,rounded corners=1pt] (9.8,0.9) rectangle (11.7,1.5) node[midway] {\tiny 14};
\draw[fill=red!20,rounded corners=1pt] (11.7,0.9) rectangle (13.3,1.5) node[midway] {\tiny 18};
\draw[fill=red!20,rounded corners=1pt] (13.3,0.9) rectangle (14.3,1.5) node[midway] {\tiny 23};
\draw[fill=red!20,rounded corners=1pt] (14.3,0.9) rectangle (16.1,1.5) node[midway] {\tiny 2};
\draw[fill=red!20,rounded corners=1pt] (16.1,0.9) rectangle (19.7,1.5) node[midway] {\tiny 6};

\end{tikzpicture}%
}
\end{center}

\end{tcolorbox}

\caption{Example of an optimal schedule generated by CP-SAT for a representative DAG instance, showing task assignments, timing, and processor-level execution.}
\label{fig:cp_sat_output}

\end{figure}

\vspace{0.5em}

\noindent

\textbf{Example:} A timeline comparison between HEFT, BO-recovered scheduling, and CP-SAT optimal scheduling was shown in Fig.~5. To concretely illustrate the structure of an optimal schedule, a representative DAG instance was considered. The schedule was obtained using a CP-SAT solver under precedence and resource constraints. This visualization highlighted how tasks were interleaved across processors while respecting dependencies. Parallel execution was exploited whenever tasks were independent, and idle time was minimized, leading to an optimal makespan of $72$. The priority function defined in Section III was used to guide scheduling decisions. This function assigned a priority score to each task based on its structural role in the graph. Scheduling decisions were then produced using list scheduling, where the highest-priority ready task was selected at each step.

    \begin{figure*}[t]
    \centering
    \resizebox{\textwidth}{!}{%
    \begin{tikzpicture}[
        font=\small,
        >=Latex,
        every node/.style={align=center},
        box1/.style={
            draw,
            rounded corners=4pt,
            thick,
            minimum width=3.35cm,
            minimum height=1.25cm,
            fill=blue!14
        },
        box2/.style={
            draw,
            rounded corners=4pt,
            thick,
            minimum width=3.45cm,
            minimum height=1.25cm,
            fill=orange!18
        },
        box3/.style={
            draw,
            rounded corners=4pt,
            thick,
            minimum width=3.45cm,
            minimum height=1.25cm,
            fill=green!16
        },
        box4/.style={
            draw,
            rounded corners=4pt,
            thick,
            minimum width=3.65cm,
            minimum height=1.25cm,
            fill=purple!14
        },
        box5/.style={
            draw,
            rounded corners=4pt,
            thick,
            minimum width=3.85cm,
            minimum height=1.25cm,
            fill=red!14
        },
        aux/.style={
            draw,
            rounded corners=4pt,
            thick,
            minimum width=3.2cm,
            minimum height=1.05cm,
            fill=yellow!18
        },
        outbox/.style={
            draw,
            rounded corners=4pt,
            thick,
            minimum width=3.65cm,
            minimum height=1.25cm,
            fill=cyan!16
        },
        group/.style={
            draw,
            dashed,
            rounded corners=6pt,
            thick,
            inner sep=8pt
        },
        arr/.style={->, thick},
        fb/.style={->, thick, dashed}
    ]
    
    % ===================== TOP ROW =====================
    \node[box1] (dag) at (0,0) {%
    \textbf{Benchmark DAG Corpus}\\
    Synthetic DAGs (ER, BA, WS)\\
    and real kernels
    };
    
    \node[box2] (parse) at (4.7,0) {%
    \textbf{Graph Parsing \& Encoding}\\
    Task weights, dependencies,\\
    communication costs
    };
    
    \node[box3] (cpsat) at (9.5,0) {%
    \textbf{CP-SAT Supervision}\\
    Offline optimal schedules\\
    for each DAG
    };
    
    \node[box4] (bo) at (14.5,0) {%
    \textbf{Bayesian Optimization}\\
    Recover priority weights $\theta^\ast$\\
    from optimal behavior
    };
    
    \node[box5] (dataset) at (19.8,0) {%
    \textbf{Scheduling Knowledge Base}\\
    $(G,\Phi(G),m(G),\theta^\ast,\pi^\ast)$
    };
    
    % ===================== BOTTOM ROW =====================
    \node[aux] (motif) at (5.8,-3.4) {%
    \textbf{Motif Analysis}\\
    Fork-join, fan-out, fan-in,\\
    reconvergent patterns
    };
    
    \node[box4] (llm) at (12.4,-3.4) {%
    \textbf{LLM Fine-Tuning / Policy Learning}\\
    Learn graph-to-heuristic mapping\\
    from structural examples
    };
    
    \node[aux] (rl) at (12.4,-5.7) {%
    \textbf{RL Refinement}\\
    Reward from makespan gap
    };
    
    \node[outbox] (infer) at (18.4,-3.4) {%
    \textbf{Inference on Unseen DAGs}\\
    Predict $\hat{\theta}$ or select\\
    the next ready task directly
    };
    
    \node[outbox] (sched) at (23.5,-3.4) {%
    \textbf{Fast List Scheduling}\\
    Near-optimal schedule\\
    with low runtime overhead
    };
    
    % ===================== MAIN TOP ARROWS =====================
    \draw[arr] (dag.east) -- (parse.west);
    \draw[arr] (parse.east) -- (cpsat.west);
    \draw[arr] (cpsat.east) -- (bo.west);
    \draw[arr] (bo.east) -- (dataset.west);
    
    % ===================== CLEAN LOWER CONNECTIONS =====================
    % parse -> motif
    \draw[arr]
        ([xshift=-0.2cm]parse.south)
        .. controls +(0,-1.2) and +(-1.2,0)
        .. (motif.north);
    
    % motif -> llm
    \draw[arr] (motif.east) -- (llm.west);
    
    % dataset -> llm routed from above, then down into llm
    \draw[arr] (dataset.south) -- ++(0,-1.1) -| ([yshift=8pt]llm.north);
    
    % llm -> infer
    \draw[arr] (llm.east) -- (infer.west);
    
    % infer -> sched
    \draw[arr] (infer.east) -- (sched.west);
    
    % ===================== RL FEEDBACK =====================
    % rl -> llm (straight up)
    \draw[fb] (rl.north) -- (llm.south);
    
    % infer -> rl reward loop routed OUTSIDE boxes
    \draw[fb] (infer.south) -- ++(0,-0.9) -| ([xshift=0.25cm]rl.east);
    
    % ===================== GROUP BOXES =====================
    \node[group, fit=(dag)(parse)(cpsat)(bo)(dataset),
          label={[font=\bfseries]above:Offline supervision and heuristic recovery}] {};
    
    \node[group, fit=(motif)(llm)(rl)(infer)(sched),
          label={[font=\bfseries]below:Learning-based generalization and deployment}] {};
    
    \end{tikzpicture}%
    }
    \caption{Overview of the MoSAIC framework. A corpus of synthetic and real benchmark DAGs is converted into structural task-graph representations containing node costs, dependencies, communication information, and motif descriptors. Each DAG is solved offline using CP-SAT to obtain optimal schedules, which provide supervision for Bayesian Optimization to recover interpretable priority weights $\theta^\ast$. The resulting knowledge base links graph structure, motif context, optimal scheduling traces, and recovered heuristics. A learning-based model, including an LLM refined through reinforcement learning, uses this supervision to predict scheduling behavior for unseen DAGs. At inference time, the predicted heuristic or next-task decision is used within a lightweight list scheduler to generate near-optimal schedules with much lower runtime than repeated exact optimization.}
    \label{fig:mosaic_framework}
    \end{figure*}
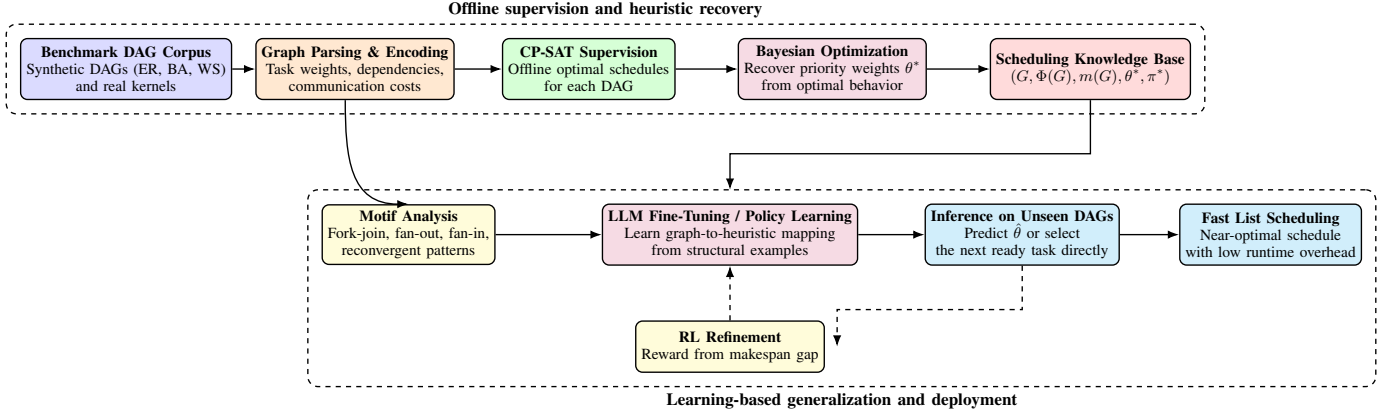

The priority parameters $\theta$ were recovered using Bayesian Optimization as described earlier. Through this process, an interpretable priority rule was recovered for each DAG. Across multiple graphs, this produced a dataset linking graph structure, optimal scheduling behavior, and effective priority functions.

\begin{figure}[t]
\centering
\resizebox{\columnwidth}{!}{%
\begin{tikzpicture}[
    x=0.12cm,
    y=0.62cm,
    >=Latex,
    prolabel/.style={anchor=east, font=\small},
    task/.style={
        draw,
        rounded corners=1pt,
        minimum height=0.42cm,
        font=\scriptsize,
        align=center
    }
]

% ================================
% HEFT
% ================================
\node[anchor=west, font=\bfseries\footnotesize] at (0,6.35) {HEFT};
\node[prolabel] at (-1.0,5.45) {P0};
\node[prolabel] at (-1.0,4.75) {P1};

\draw[->] (0,5.75) -- (55,5.75);
\foreach \x in {0,10,20,30,40,50}
    \node[font=\tiny] at (\x,6.0) {\x};

% HEFT P0
\draw[task, fill=blue!20]   (0,5.2) rectangle (7,5.6) node[pos=.5]{T1};
\draw[task, fill=green!20]  (9,5.2) rectangle (17,5.6) node[pos=.5]{T3};
\draw[task, fill=orange!25] (19,5.2) rectangle (28,5.6) node[pos=.5]{T6};
\draw[task, fill=purple!20] (31,5.2) rectangle (40,5.6) node[pos=.5]{T8};
\draw[task, fill=red!20]    (44,5.2) rectangle (53,5.6) node[pos=.5]{T10};

% HEFT P1
\draw[task, fill=cyan!20]    (0,4.5) rectangle (6,4.9) node[pos=.5]{T2};
\draw[task, fill=yellow!30]  (8,4.5) rectangle (14,4.9) node[pos=.5]{T4};
\draw[task, fill=teal!20]    (16,4.5) rectangle (24,4.9) node[pos=.5]{T5};
\draw[task, fill=magenta!20] (27,4.5) rectangle (36,4.9) node[pos=.5]{T7};
\draw[task, fill=brown!20]   (39,4.5) rectangle (48,4.9) node[pos=.5]{T9};

\draw[dashed] (53,4.35) -- (53,5.75);
\node[font=\scriptsize] at (50.5,4.1) {$C_{\max}=53$};

% ================================
% BO-Recovered
% ================================
\node[anchor=west, font=\bfseries\footnotesize] at (0,3.85) {BO-Recovered};
\node[prolabel] at (-1.0,3.0) {P0};
\node[prolabel] at (-1.0,2.3) {P1};

\draw[->] (0,3.3) -- (55,3.3);
\foreach \x in {0,10,20,30,40,50}
    \node[font=\tiny] at (\x,3.55) {\x};

% BO P0
\draw[task, fill=blue!20]   (0,2.75) rectangle (7,3.15) node[pos=.5]{T1};
\draw[task, fill=green!20]  (7,2.75) rectangle (15,3.15) node[pos=.5]{T3};
\draw[task, fill=orange!25] (15,2.75) rectangle (24,3.15) node[pos=.5]{T6};
\draw[task, fill=purple!20] (26,2.75) rectangle (35,3.15) node[pos=.5]{T8};
\draw[task, fill=red!20]    (37,2.75) rectangle (46,3.15) node[pos=.5]{T10};

% BO P1
\draw[task, fill=cyan!20]    (0,2.05) rectangle (6,2.45) node[pos=.5]{T2};
\draw[task, fill=yellow!30]  (6,2.05) rectangle (12,2.45) node[pos=.5]{T4};
\draw[task, fill=teal!20]    (12,2.05) rectangle (20,2.45) node[pos=.5]{T5};
\draw[task, fill=magenta!20] (20,2.05) rectangle (29,2.45) node[pos=.5]{T7};
\draw[task, fill=brown!20]   (31,2.05) rectangle (40,2.45) node[pos=.5]{T9};

\draw[dashed] (46,1.9) -- (46,3.3);
\node[font=\scriptsize] at (43.5,1.65) {$C_{\max}=46$};

% ================================
% CP-SAT Optimal
% ================================
\node[anchor=west, font=\bfseries\footnotesize] at (0,1.35) {CP-SAT Optimal};
\node[prolabel] at (-1.0,0.55) {P0};
\node[prolabel] at (-1.0,-0.15) {P1};

\draw[->] (0,0.85) -- (55,0.85);
\foreach \x in {0,10,20,30,40,50}
    \node[font=\tiny] at (\x,1.1) {\x};

% OPT P0
\draw[task, fill=blue!20]   (0,0.3) rectangle (7,0.7) node[pos=.5]{T1};
\draw[task, fill=green!20]  (7,0.3) rectangle (15,0.7) node[pos=.5]{T3};
\draw[task, fill=orange!25] (15,0.3) rectangle (24,0.7) node[pos=.5]{T6};
\draw[task, fill=purple!20] (24,0.3) rectangle (33,0.7) node[pos=.5]{T8};
\draw[task, fill=red!20]    (33,0.3) rectangle (42,0.7) node[pos=.5]{T10};

% OPT P1
\draw[task, fill=cyan!20]    (0,-0.4) rectangle (6,0.0) node[pos=.5]{T2};
\draw[task, fill=yellow!30]  (6,-0.4) rectangle (12,0.0) node[pos=.5]{T4};
\draw[task, fill=teal!20]    (12,-0.4) rectangle (20,0.0) node[pos=.5]{T5};
\draw[task, fill=magenta!20] (20,-0.4) rectangle (29,0.0) node[pos=.5]{T7};
\draw[task, fill=brown!20]   (29,-0.4) rectangle (38,0.0) node[pos=.5]{T9};

\draw[dashed] (42,-0.55) -- (42,0.85);
\node[font=\scriptsize] at (39.5,-0.8) {$C_{\max}=42$};

\end{tikzpicture}%
}
\vspace{-4pt}
\caption{Scheduling timeline comparison on the same DAG instance using HEFT, BO-recovered scheduling, and CP-SAT optimal scheduling. The learned policy reduces idle gaps and achieves makespan closer to the optimal solution.}
\label{fig:timeline_singlecol}
\end{figure}
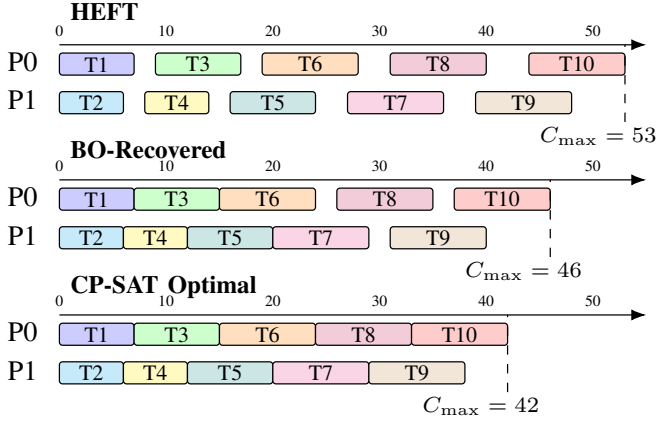
\vspace{-1pt}

\subsection{Formal Insight on Priority Function Recovery}

The recovered priority function provided a compact approximation of the decision behavior induced by optimal scheduling. Instead of explicitly reproducing optimal schedules, the method captured task-selection preferences through a low-dimensional feature representation.

At each step, the scheduler selected:
\begin{equation}
v_t^* = \arg\max_{v \in R_t} H(v; \theta)
\end{equation}

The pairwise preference between two candidate tasks was given by:
\begin{equation}
\Delta H(u, v; \theta) = H(u; \theta) - H(v; \theta)
\end{equation}

The priority function was defined as a weighted combination of structural features:
\begin{equation}
H(v; \theta) = \sum_{k=1}^{d} \theta_k \, \phi_k(v)
\end{equation}

In expanded form, using key scheduling features:
\begin{equation}
\begin{aligned}
H(v; \theta) =\;&
\theta_1 \cdot \text{rank}_u(v) +
\theta_2 \cdot \text{depth}(v) \\
&+ \theta_3 \cdot \text{fanout}(v) +
\theta_4 \cdot \text{indegree}(v) \\
&+ \theta_5 \cdot \bar{c}_{out}(v)
\end{aligned}
\end{equation}

This formulation allowed scheduling behavior to be interpreted as a decision surface over structural feature space:
\begin{equation}
H(v; \theta) = \theta^\top \phi(v)
\end{equation}

As a result, the recovered heuristic approximated the ordering decisions of the optimal solver while remaining computationally efficient and interpretable.

\subsection{LLM-Based Scheduling and RL Refinement}

The constructed dataset was used to train a large language model (LLM) to perform scheduling decisions. Scheduling was formulated as a structured reasoning task over graph representations rather than as free-form text generation. While Bayesian Optimization recovered a priority function for each individual graph instance, it did not generalize to unseen graphs without re-optimization. The LLM addressed this limitation by learning a transferable mapping from graph structure to scheduling behavior across the entire training corpus.

Each training instance consisted of graph-level descriptors, task-level features, structural motifs, optimal scheduling decisions, and recovered priority functions. Motifs such as fork-join structures, linear chains, and reconvergent dependencies were explicitly included as they captured recurring local patterns that influenced scheduling behavior. This representation exposed the model not only to task attributes but also to structural context that governed scheduling decisions. During fine-tuning, the LLM learned to associate these structural signals with appropriate task-selection actions. For example, when the ready queue contained a task on the critical path with high upward rank and another task with lower downstream impact, the model learned to prioritize the former. Similarly, in fork-join regions, delaying a critical branch could increase synchronization latency, and the model learned to avoid such decisions.

Scheduling was modeled as a sequential decision process, where each step involved selecting a task from the ready set. To further improve decision quality, reinforcement learning was applied. Unlike purely reinforcement learning-based approaches, which require extensive exploration and may converge slowly, this framework leverages optimal schedules as supervision to guide policy learning, improving both sample efficiency and stability. At each step:
\begin{itemize}
    \item the \textit{state} consisted of the current partial schedule and the ready task set,
    \item the \textit{action} corresponded to selecting the next task, and
    \item the \textit{reward} reflected the impact of that decision on the final schedule quality.
\end{itemize}
The reward was defined as:
\begin{equation}
R = - \left(C_{\text{current}} - C_{\text{optimal}}\right)
\end{equation}

This reward formulation assigned higher values to decisions that kept the generated schedule closer to the optimal one. For instance, in a fork-join structure, selecting a non-critical ready task could delay the join node and increase the makespan, resulting in a lower reward. In contrast, prioritizing tasks that preserved critical-path progress yielded higher reward. Through repeated interactions, the model refined its policy and progressively aligned its decisions with those implied by optimal schedules.

We instantiated the learning-based scheduler using GPT-5.2 and adapted it via LoRA-based supervised fine-tuning \cite{ref21} followed by reinforcement learning. Training data was constructed at the decision level: each CP-SAT optimal schedule was decomposed into a sequence of next-task selections. At each step, the model received a serialized state consisting of graph descriptors, motif context, ready-queue tasks, and partial schedule information, and was trained to predict the next task chosen in the optimal trace. For example, if the ready queue contained two tasks where one lay on the critical path with higher upward rank and the other had lower downstream impact, the model learned to prioritize the former. After supervised training, reinforcement learning was applied over full scheduling episodes, where the reward was defined by the normalized makespan gap to the CP-SAT optimum.

\vspace{3pt}
\subsection{Structured Prompting View of the LLM}
%\vspace{-10pt}

The LLM was not used to generate an entire schedule in unconstrained natural language form. Instead, it solved a structured decision problem in which the current graph context, ready queue, task descriptors, and motif annotations were serialized into a consistent input format and the model selected the next task. This distinction was important because the LLM operated as a policy approximator over structured scheduling states rather than as a generic text generator. This structured formulation constrained the decision space and improved consistency across scheduling instances with varying graph sizes and topologies. By conditioning on both local task features and global structural cues, the model captured dependencies that were difficult to encode in fixed heuristic rules. As a result, the learned policy generalized across diverse DAGs without requiring per-instance re-optimization.
\vspace{-1pt}
A representative prompt instance could be organized as shown in Fig.~6.

\begin{figure}[t]
\centering

\begin{tcolorbox}[
colback=gray!5,
colframe=black,
title=LLM Scheduling Decision Context,
boxrule=0.5pt,
sharp corners,
fonttitle=\bfseries
]

\small
\textbf{Graph Features:} Family = BA, Average depth = 3.2, Motif = fork-join

\textbf{Ready Tasks:} 
$v_8$ (rank = 50, depth = 0, fanout = 4), 
$v_{21}$ (rank = 47, depth = 0, fanout = 2)

\textbf{Heuristic Context:} $\theta_{\text{rank}}$ high, $\theta_{\text{fanout}}$ negative

\textbf{Decision Objective:} Select the next task that best preserves schedule quality

\end{tcolorbox}

\caption{Example of the structured input representation provided to the LLM for scheduling decisions, combining graph-level features, ready-task descriptors, and heuristic context.}
\label{fig:llm_decision_context}

\end{figure}

This format allowed the model to reason over both numeric task descriptors and graph-structural cues. The input could also include the current partial schedule and processor availability so that the task-selection decision remained tied to the actual scheduling state. In this way, the LLM was trained to learn which structural combinations typically corresponded to good scheduling choices, instead of merely memorizing graph identities or sequence traces.

\subsection{Results on Real Benchmarks}

Table~\ref{tab:real_perf} reports the makespan obtained on the real benchmark DAGs using CP-SAT \cite{ref04}, HEFT \cite{ref02, ref23}, the recovered BO-based priority rule, and the LLM+RL policy.

\begin{table}[t]
\centering
\caption{Makespan comparison on real benchmarks. CP-SAT provides the optimal reference and HEFT is the baseline. Learned methods reduce the gap, with LLM+RL closest to optimal.}
\label{tab:real_perf}
\setlength{\tabcolsep}{6pt}
\renewcommand{\arraystretch}{1.15}
\fbox{%
\begin{tabular}{@{\hspace{6pt}}lrrrr@{\hspace{6pt}}}
\toprule
\textbf{Kernel} & \textbf{CP-SAT} & \textbf{HEFT} & \textbf{BO} & \textbf{LLM+RL} \\
\midrule
AES      & 103 & 108 & 106 & 105 \\
ATAX     & 132 & 138 & 136 & 134 \\
BICG     & 122 & 127 & 126 & 124 \\
CONV2D   & 172 & 180 & 178 & 176 \\
FFT      & 169 & 175 & 174 & 172 \\
GEMM     & 107 & 112 & 110 & 109 \\
STENCIL  & 140 & 145 & 145$^{*}$ & 143 \\
SYRK     & 115 & 128 & 118 & 117 \\
\bottomrule
\multicolumn{5}{@{\hspace{6pt}}l}{\footnotesize $^{*}$BO matches HEFT on STENCIL}
\end{tabular}%
}
\end{table}

The real benchmark results showed that HEFT provided a strong baseline for traditional heuristic scheduling, but its fixed ranking mechanism was less effective on kernels with more complex dependency interactions. The BO-based policy reduced the gap by adapting the feature weights to graph-specific structure. The LLM+RL model moved further in that direction by learning decision behavior across multiple graphs instead of relying only on one recovered rule per instance. The largest improvements appeared in kernels such as SYRK and CONV2D, where dependency structure created stronger interactions between task ordering, synchronization, and processor utilization. In such cases, static heuristics were less flexible because they applied the same ranking logic across all graphs, while the learned methods adapted to the structural characteristics of each instance. Although evaluated on dual-processor settings, these gains stemmed from structure-aware prioritization and were not tied to processor count, enabling generalization to multi-processor scenarios.

\begin{table}[t]
\centering
\caption{Observations from Real Benchmark Results}
\label{tab:real_obs}
\setlength{\tabcolsep}{6pt}
\renewcommand{\arraystretch}{1.15}
\fbox{%
\begin{tabular}{@{\hspace{6pt}}p{0.26\linewidth} p{0.60\linewidth}@{\hspace{6pt}}}
\toprule
\textbf{Aspect} & \textbf{Observation} \\
\midrule
HEFT behavior & HEFT produces competitive schedules on simpler kernels but shows larger deviation from the optimal schedule on graphs with stronger structural irregularity. \\[4pt]
BO recovery & The BO-based priority rule reduces the makespan gap by adapting task importance to graph-level and task-level features. \\[4pt]
LLM+RL behavior & The LLM+RL policy improves over BO in most cases by learning decision-level patterns from multiple graphs and refining them through reward feedback. \\[4pt]
Most affected kernels & SYRK and CONV2D show the clearest benefit from learned scheduling, indicating stronger sensitivity to local dependency structure. \\
\bottomrule
\end{tabular}%
}
\end{table}

To broaden the comparison beyond HEFT, additional classical baselines were also considered. These included CPOP, DLS, and a forced scheduling strategy. The purpose of this comparison was not only to compare absolute makespan values, but also to verify that the gains of the recovered and learned policies were not limited to outperforming a single classical heuristic.

\begin{table}[t]
\centering
\caption{Representative Comparison with Additional Baselines}
\label{tab:extra_baselines}
\scriptsize
\setlength{\tabcolsep}{3pt}
\renewcommand{\arraystretch}{1.08}
\fbox{%
\begin{tabular*}{0.96\columnwidth}{@{\extracolsep{\fill}}lcccc@{}}
\toprule
\textbf{Method} & \textbf{Syn.} & \textbf{Real} & \textbf{Interp.} & \textbf{Adapt.} \\
                & \textbf{Gap (\%)} & \textbf{Gap (\%)} & & \\
\midrule
Force-Directed Scheduling~\cite{ref24} & 6.4 & 5.8 & Low    & Low    \\
CPOP~\cite{ref02, ref23}                      & 5.9 & 5.1 & Medium & Low    \\
DLS~\cite{ref23}                       & 5.4 & 4.9 & Medium & Medium \\
HEFT~\cite{ref02}                      & 5.6 & 4.7 & High   & Low    \\
BO-Recovered (Ours)                    & 2.8 & 2.2 & High   & High   \\
LLM+RL (Ours)                          & 2.1 & 1.4 & Medium & High   \\
\bottomrule
\end{tabular*}%
}
\end{table}

The expanded comparison indicated that the improvement was not tied to any single baseline. These qualitative observations are summarized in Table~II. Classical heuristics remained effective when the graph structure aligned with their ranking assumptions, but their fixed decision logic became less reliable as structural irregularity increased. The recovered BO rule and the LLM+RL policy preserved the computational simplicity of priority-based scheduling while adapting more effectively to graph-specific structure.

\subsection{Results on Synthetic Benchmarks}

To study structural generalization, the methods were also compared on synthetic DAG families. Table~\ref{tab:synth_gap} reported representative average gap values relative to the optimal schedule. The synthetic results revealed a similar pattern across all graph families. HEFT performed reasonably well, but the learned methods remained closer to the optimal schedule. BA graphs showed the highest gap values overall because their hub-dominated structure created stronger scheduling imbalance and more difficult ordering choices. ER and WS graphs were comparatively more regular, although they still benefited from learned scheduling behavior.

\begin{table}[t]
\centering
\caption{Average Gap Comparison on Synthetic DAGs (\%)}
\label{tab:synth_gap}
\setlength{\tabcolsep}{6pt}
\renewcommand{\arraystretch}{1.15}
\fbox{%
\begin{tabular}{@{\hspace{6pt}}lccc@{\hspace{6pt}}}
\toprule
\textbf{Graph Type} & \textbf{HEFT} & \textbf{BO} & \textbf{LLM+RL} \\
\midrule
ER & 5.2 & 2.8 & 2.1 \\
BA & 6.0 & 3.1 & 2.4 \\
WS & 5.5 & 2.6 & 2.0 \\
\bottomrule
\end{tabular}%
}
\end{table}

\subsection{Generalization Across Graph Families}

A stronger test of learning quality was whether a scheduling policy trained on one set of structures remained effective on structurally different DAGs. For this purpose, cross-family evaluation was considered, where the training set was drawn from one or two synthetic families and testing was performed on a different family or on real benchmark graphs.

\begin{table}[t]
\centering
\caption{Representative Generalization Results}
\label{tab:generalization}
\setlength{\tabcolsep}{6pt}
\renewcommand{\arraystretch}{1.15}
\fbox{%
\begin{tabular}{@{\hspace{6pt}}lcc@{\hspace{6pt}}}
\toprule
\textbf{Train Set} & \textbf{Test Set} & \textbf{Avg. Gap (\%)} \\
\midrule
ER + BA    & WS              & 2.5 \\
ER + WS    & BA              & 2.9 \\
Synthetic  & Real Benchmarks & 2.3 \\
\bottomrule
\end{tabular}%
}
\end{table}

These results indicated that the learned policy was not merely memorizing training graphs. Performance degraded only moderately when the test distribution changed, which suggested that the model was learning reusable structural decision patterns rather than graph-specific sequences. This was particularly important for real deployment settings, where the policy had to operate on unseen task graphs.

\subsection{Ablation Study}

To isolate the role of each component, an ablation study was conducted by progressively removing motifs, Bayesian Optimization, and reinforcement learning from the full scheduling pipeline. The comparison focused on makespan gap relative to the optimal schedule. The ablation showed that each stage contributed a distinct role. The quantitative results of this ablation study are summarized in Table~VI. Motifs injected structural context beyond flat task attributes, BO recovered a useful priority surface from optimal supervision, and RL sharpened local decisions by connecting them to long-horizon schedule quality. The largest degradation appeared when BO was removed, which indicated that recovering task-priority structure from optimal schedules was central to the overall approach.

\begin{table}[t]
\centering
\caption{Ablation Study of the Scheduling Pipeline}
\label{tab:ablation}
\setlength{\tabcolsep}{6pt}
\renewcommand{\arraystretch}{1.15}
\fbox{%
\begin{tabular}{@{\hspace{6pt}}lc@{\hspace{6pt}}}
\toprule
\textbf{Variant} & \textbf{Avg. Gap (\%)} \\
\midrule
Full framework           & 2.1 \\
Without motifs           & 3.0 \\
Without BO (random $\theta$) & 5.5 \\
Without RL               & 2.8 \\
\bottomrule
\end{tabular}%
}
\end{table}

\subsection{Scalability Analysis}
While exact optimization was useful for supervision, it was 
impractical for large graphs. The scaling behavior of HEFT, BO-recovered scheduling, and LLM+RL across increasing graph sizes was as shown in Fig.~7. The timing results in 
Table~\ref{tab:scalability} showed a clear boundary beyond 
150 nodes for CP-SAT, whereas the learned inference policy 
and recovered rule remained in low-latency ranges, making 
them suitable for online scheduling. For large DFGs, the 
LLM operated on a fixed-length serialized ready set rather 
than the full graph, resulting in scaling proportional to 
the ready set size.

With respect to processor count, the priority function was 
not restricted to two processors and generalized to $m$ 
processors through standard list scheduling. The learned 
policy $\theta^*$ could be reused without retraining. The main 
challenge in extending to higher processor counts lay in 
obtaining optimal CP-SAT supervision within practical time 
limits.

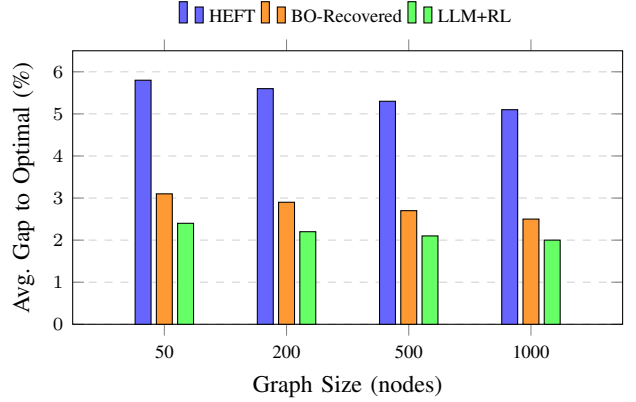
\begin{figure}[t]
\centering
\begin{tikzpicture}
\begin{axis}[
    width=\columnwidth,
    height=5.2cm,
    ybar,
    bar width=6pt,
    enlarge x limits=0.25,
    ylabel={Avg. Gap to Optimal (\%)},
    xlabel={Graph Size (nodes)},
    symbolic x coords={50,200,500,1000},
    xtick=data,
    ymin=0, ymax=6.5,
    ytick={0,1,2,3,4,5,6},
    ymajorgrids=true,
    xmajorgrids=false,
    grid style={dashed, gray!35},
    legend style={
        at={(0.5,1.05)},
        anchor=south,
        legend columns=3,
        font=\scriptsize,
        draw=none
    },
    tick label style={font=\scriptsize},
    label style={font=\small}
]

% HEFT
\addplot[
    fill=blue!60,
    draw=black
] coordinates {
    (50,5.8)
    (200,5.6)
    (500,5.3)
    (1000,5.1)
};

% BO
\addplot[
    fill=orange!80,
    draw=black
] coordinates {
    (50,3.1)
    (200,2.9)
    (500,2.7)
    (1000,2.5)
};

% LLM+RL
\addplot[
    fill=green!60,
    draw=black
] coordinates {
    (50,2.4)
    (200,2.2)
    (500,2.1)
    (1000,2.0)
};

\legend{HEFT, BO-Recovered, LLM+RL}

\end{axis}
\end{tikzpicture}

\vspace{-4pt}
\caption{Performance scaling across increasing graph sizes. Learned scheduling policies (BO and LLM+RL) consistently achieve lower optimality gaps than HEFT across all graph sizes.}
\label{fig:gap_bar}
\vspace{-6pt}
\end{figure}

\begin{table}[t]
\centering
\caption{Representative Runtime Scalability Comparison}
\label{tab:scalability}
\setlength{\tabcolsep}{6pt}
\renewcommand{\arraystretch}{1.15}
\fbox{%
\begin{tabular}{@{\hspace{6pt}}lccc@{\hspace{6pt}}}
\toprule
\textbf{Method} & \textbf{200 Nodes} & \textbf{500 Nodes} & \textbf{1000 Nodes} \\
\midrule
CP-SAT            & Timeout & Timeout & Timeout \\
HEFT              & 10 ms   & 14 ms   & 21 ms   \\
BO-Recovered Rule & 16 ms   & 23 ms   & 35 ms   \\
LLM+RL            & 20 ms   & 31 ms   & 48 ms   \\
\bottomrule
\end{tabular}%
}
\end{table}

\subsection{LLM Scheduling Accuracy}

In addition to final makespan, the LLM was evaluated at the decision level by measuring how often its selected task matched the intended scheduling action in the training supervision. Table~\ref{tab:llm_acc} reported representative accuracy values for zero-shot, fine-tuned, and RL-refined settings. These values showed that zero-shot reasoning alone was not sufficient for capturing the structure of scheduling decisions. A structured summary of these observations is provided in Table~IX. Fine-tuning improved the model by exposing it to graph features, motifs, and optimal scheduling examples. Reinforcement learning further sharpened the policy by linking each decision to its long-term effect on the schedule.

\subsection{Interpretability of Learned Priority Weights}

A major advantage of the recovered priority function was that it remained interpretable even after learning. The coefficient values directly indicated which structural properties were emphasized by the scheduling policy on different graph types. This motif-conditioned variation in scheduling behavior was illustrated in Fig.~8.

This perspective at the coefficient level also shed light on why this method was better than fixed heuristics. Instead of applying a single principle of ranking to all graphs, the priority surface learned changed depending on the structure and dependency of each graph. Thus, this method remained interpretable while also adapting to the structure of the graph. The novelty of MoSAIC lies in converting optimal schedules, obtained through expensive exact optimization, into reusable and interpretable scheduling policies. Instead of directly using optimal solutions, Bayesian Optimization distills their behavior into compact priority functions, which then serve as structured supervision for learning-based reasoning. This establishes a systematic bridge between exact optimization and generalizable heuristic design, rather than treating them as independent approaches.

From a systems perspective, MoSAIC showed that optimal scheduling in heterogeneous DAGs was structure dependent. Different graph families induced different priority behaviors: BA graphs emphasized critical-path urgency and avoided premature hub expansion, WS graphs required synchronization-aware decisions, and ER graphs balanced urgency with topological progress. Static heuristics such as HEFT failed because they applied a single ranking rule across all graphs, whereas MoSAIC adapted to structural variation.

% in preamble

% ================= FIGURE =================
\begin{figure}[t]
\centering
\footnotesize

\begin{minipage}{0.95\columnwidth}
\centering

% ---------- Panel A ----------
\begin{tcolorbox}[
colback=blue!6,
colframe=blue!45!black,
boxrule=0.5pt,
arc=2pt,
left=4pt,right=4pt,top=4pt,bottom=4pt,
width=\linewidth
]
\centering
\begin{tikzpicture}[scale=0.78, every node/.style={font=\scriptsize}]
% nodes
\node[circle, draw, fill=blue!12, minimum size=5.5mm] (a1) at (0,1.6) {$v_1$};
\node[circle, draw, fill=blue!12, minimum size=5.5mm] (a2) at (1.4,2.2) {$v_2$};
\node[circle, draw, fill=blue!12, minimum size=5.5mm] (a3) at (1.4,1.0) {$v_3$};
\node[circle, draw, fill=blue!12, minimum size=5.5mm] (a4) at (1.4,-0.2) {$v_4$};
\node[circle, draw, fill=blue!18, minimum size=5.5mm] (a5) at (3.0,1.0) {$v_5$};

% edges
\draw[-{Latex[length=2mm]}, thick] (a1) -- (a2);
\draw[-{Latex[length=2mm]}, thick] (a1) -- (a3);
\draw[-{Latex[length=2mm]}, thick] (a1) -- (a4);
\draw[-{Latex[length=2mm]}, thick] (a2) -- (a5);
\draw[-{Latex[length=2mm]}, thick] (a3) -- (a5);
\draw[-{Latex[length=2mm]}, thick] (a4) -- (a5);

% title + text
\node[font=\scriptsize\bfseries] at (1.5,-1.0) {(a) BA-like hub-dominated motif};
\node[align=center, text width=6.8cm, font=\scriptsize] at (1.5,-1.8)
{High $\theta_{\mathrm{rank}}$, negative $\theta_{\mathrm{fanout}}$.\\
Critical-path urgency is emphasized while premature fanout expansion is discouraged.};
\end{tikzpicture}
\end{tcolorbox}

\vspace{1.5mm}

% ---------- Panel B ----------
\begin{tcolorbox}[
colback=green!6,
colframe=green!45!black,
boxrule=0.5pt,
arc=2pt,
left=4pt,right=4pt,top=4pt,bottom=4pt,
width=\linewidth
]
\centering
\begin{tikzpicture}[scale=0.78, every node/.style={font=\scriptsize}]
% nodes
\node[circle, draw, fill=green!12, minimum size=5.5mm] (b1) at (0,1.2) {$v_1$};
\node[circle, draw, fill=green!12, minimum size=5.5mm] (b2) at (1.3,2.0) {$v_2$};
\node[circle, draw, fill=green!12, minimum size=5.5mm] (b3) at (1.3,0.4) {$v_3$};
\node[circle, draw, fill=green!12, minimum size=5.5mm] (b4) at (2.8,1.2) {$v_4$};
\node[circle, draw, fill=green!18, minimum size=5.5mm] (b5) at (4.2,1.2) {$v_5$};

% edges
\draw[-{Latex[length=2mm]}, thick] (b1) -- (b2);
\draw[-{Latex[length=2mm]}, thick] (b1) -- (b3);
\draw[-{Latex[length=2mm]}, thick] (b2) -- (b4);
\draw[-{Latex[length=2mm]}, thick] (b3) -- (b4);
\draw[-{Latex[length=2mm]}, thick] (b4) -- (b5);

% title + text
\node[font=\scriptsize\bfseries] at (2.1,-0.9) {(b) ER-like mixed motif};
\node[align=center, text width=6.8cm, font=\scriptsize] at (2.1,-1.7)
{Balanced $\theta_{\mathrm{rank}}$ and $\theta_{\mathrm{depth}}$.\\
The policy trades off urgency and topological progress more uniformly.};
\end{tikzpicture}
\end{tcolorbox}

\vspace{1.5mm}

% ---------- Panel C ----------
\begin{tcolorbox}[
colback=orange!8,
colframe=orange!60!black,
boxrule=0.5pt,
arc=2pt,
left=4pt,right=4pt,top=4pt,bottom=4pt,
width=\linewidth
]
\centering
\begin{tikzpicture}[scale=0.78, every node/.style={font=\scriptsize}]
% nodes
\node[circle, draw, fill=orange!12, minimum size=5.5mm] (c1) at (0,1.8) {$v_1$};
\node[circle, draw, fill=orange!12, minimum size=5.5mm] (c2) at (0,0.4) {$v_2$};
\node[circle, draw, fill=orange!12, minimum size=5.5mm] (c3) at (1.5,1.8) {$v_3$};
\node[circle, draw, fill=orange!12, minimum size=5.5mm] (c4) at (1.5,0.4) {$v_4$};
\node[circle, draw, fill=orange!18, minimum size=5.5mm] (c5) at (3.1,1.1) {$v_5$};

% edges
\draw[-{Latex[length=2mm]}, thick] (c1) -- (c3);
\draw[-{Latex[length=2mm]}, thick] (c2) -- (c4);
\draw[-{Latex[length=2mm]}, thick] (c3) -- (c5);
\draw[-{Latex[length=2mm]}, thick] (c4) -- (c5);

% title + text
\node[font=\scriptsize\bfseries] at (1.6,-0.9) {(c) WS-like clustered merge motif};
\node[align=center, text width=6.8cm, font=\scriptsize] at (1.6,-1.7)
{Positive $\theta_{\mathrm{depth}}$ and $\theta_{\mathrm{indegree}}$.\\
The policy reacts more strongly to synchronization-heavy merge structure.};
\end{tikzpicture}
\end{tcolorbox}

\end{minipage}

\vspace{1mm}
\caption{Motif-conditioned interpretation of learned priority behavior across graph families.
Different local structures induce different scheduling preferences.
BA-like hub-dominated regions emphasize critical-path urgency while discouraging premature fanout expansion.
ER-like regions balance urgency and topological progress.
WS-like clustered and merge-heavy regions assign stronger importance to depth and indegree to reduce synchronization delay.
This illustrates that effective scheduling behavior is graph-structure dependent rather than governed by a single fixed priority rule.}
\label{fig:motif_priority_interpretation}
\end{figure}
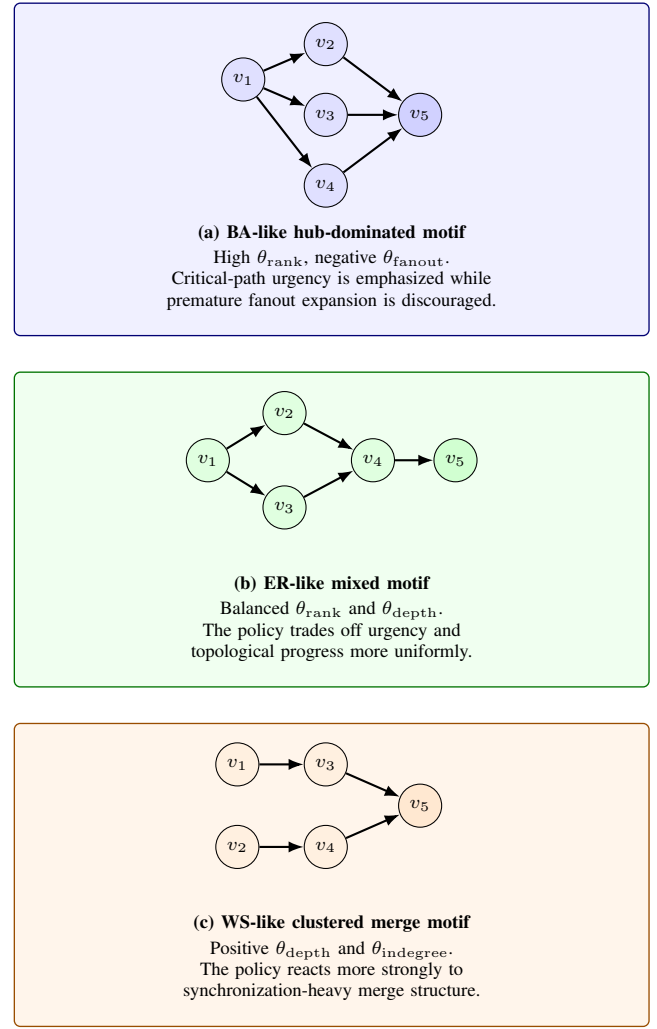

\begin{table}[t]
\centering
\caption{LLM scheduling decision accuracy (mean $\pm$ std) relative to the optimal schedule.}
\label{tab:llm_acc}
\setlength{\tabcolsep}{6pt}
\renewcommand{\arraystretch}{1.15}
\fbox{%
\begin{tabular}{@{\hspace{6pt}}lc@{\hspace{6pt}}}
\toprule
\textbf{Method} & \textbf{Accuracy} \\
\midrule
Zero-shot LLM  & 46.3 $\pm$ 4.8\% \\
Fine-tuned LLM & 67.5 $\pm$ 3.9\% \\
LLM + RL       & \textbf{80.8 $\pm$ 2.7\%} \\
\bottomrule
\end{tabular}%
}
\end{table}

\begin{table}[t]
\centering
\caption{Observations from LLM Scheduling Accuracy}
\label{tab:llm_obs}
\setlength{\tabcolsep}{6pt}
\renewcommand{\arraystretch}{1.15}
\fbox{%
\begin{tabular}{@{\hspace{6pt}}p{0.26\linewidth} p{0.60\linewidth}@{\hspace{6pt}}}
\toprule
\textbf{Aspect} & \textbf{Observation} \\
\midrule
Zero-shot limitation & A general-purpose LLM without scheduling-specific training captures only a limited portion of the decision logic. \\[4pt]
Effect of fine-tuning & Fine-tuning improves the model by connecting structural graph descriptors to scheduling choices. \\[4pt]
Effect of RL & RL refines the policy by rewarding choices that lead to schedules closer to the optimal one over multiple steps. \\[4pt]
Decision quality & Higher decision accuracy aligns with improved final makespan, indicating that local action quality influences global schedule quality. \\
\bottomrule
\end{tabular}%
}
\end{table}

Beyond average performance, statistical reporting provided a clearer picture of consistency. Mean and standard deviation described the spread of makespan gap values across graph instances, while win/loss counts indicated how frequently one method outperformed another. The statistical comparison against HEFT is reported in Table~X. This view showed that the gains were not confined to a small number of outlier instances. The learned methods reduced both average gap and variability, while also winning against HEFT on a substantial majority of test graphs.

\begin{table}[t]
\centering
\caption{Statistical Summary Against HEFT}
\label{tab:stats}
\setlength{\tabcolsep}{6pt}
\renewcommand{\arraystretch}{1.15}
\fbox{%
\begin{tabular}{@{\hspace{6pt}}lcc@{\hspace{6pt}}}
\toprule
\textbf{Method} & \textbf{Gap (\%)} & \textbf{Win Rate vs. HEFT} \\
\midrule
HEFT              & $5.6 \pm 1.8$ & --  \\
BO-Recovered Rule & $2.8 \pm 1.2$ & 71\% \\
LLM+RL            & $2.1 \pm 1.0$ & 78\% \\
\bottomrule
\end{tabular}%
}
\end{table}

\subsection{Error Analysis}

Failure analysis was useful for understanding where learned scheduling still struggled. Although the proposed framework improved average schedule quality, some graph regimes remained difficult. These cases clarified that the learned framework was most beneficial when the graph exhibited meaningful structural choice. Representative error cases and their characteristics are summarized in Table XI. When the schedule was already tightly constrained by topology, the margin for improvement naturally decreased.

Across synthetic and real benchmarks, the learned scheduling policies consistently outperformed classical heuristics by adapting to structural variations in the DAG. In hub-dominated graphs (e.g., BA), prioritizing critical-path tasks yielded larger gains, while in more regular or low-variance structures (e.g., STENCIL), improvements were limited as baseline heuristics already aligned with near-optimal decisions. Additionally, the reinforcement learning stage provided consistent refinement over supervised policies by improving long-horizon scheduling decisions, particularly in graphs with higher communication and synchronization complexity.

\begin{table}[t]
\centering
\caption{Representative Error Analysis}
\label{tab:error_analysis}
\setlength{\tabcolsep}{6pt}
\renewcommand{\arraystretch}{1.15}
\fbox{%
\begin{tabular}{@{\hspace{6pt}}p{0.26\linewidth} p{0.60\linewidth}@{\hspace{6pt}}}
\toprule
\textbf{Case} & \textbf{Observation} \\
\midrule
Dense DAGs & Limited scheduling freedom reduces the advantage of learned decision policies because precedence constraints dominate most choices. \\[4pt]
Chain-like graphs & Gains over HEFT are small because the ready set is narrow and the schedule is already strongly constrained. \\[4pt]
High communication cost cases & Errors become more likely when processor assignment and task ordering interact strongly through communication penalties. \\[4pt]
Low parallelism regions & There are fewer opportunities to exploit structural preferences, reducing the benefit of adaptive prioritization. \\
\bottomrule
\end{tabular}%
}
\end{table}

\subsection{Limitations \& Summary}
The present framework had several limitations. First, it depended 
on offline exact or near-exact supervision, which required solving 
smaller graphs with CP-SAT before reusable policies could be learned. 
Second, the current formulation was specialized to DAG scheduling 
and did not directly address cyclic or dynamically evolving task 
graphs. Third, the quality of the recovered heuristic depended on 
the expressiveness of the chosen structural features and motif 
descriptors. If important scheduling cues were absent from the 
feature representation, part of the solver's decision logic could 
remain unrecovered. Fourth, the current evaluation targeted a 
dual-processor setup; extending to architectures with non-uniform 
memory latencies or domain-specific accelerators such as FPGAs 
and TPUs was a natural next step, as the priority function 
formulation was not tied to any specific hardware model. The experimental results showed that optimal schedules could be used 
to derive interpretable and adaptive scheduling policies. The 
combination of exact optimization, feature-based heuristic 
recovery, and learning-based decision models provided a framework 
for near-optimal scheduling with lower runtime complexity than 
repeated exact solving. The LLM and RL components extended this 
framework by learning reusable scheduling behavior from graph 
structure, motifs, optimal schedules, and recovered priority 
functions. The additional analyses further showed that the framework 
benefited from each of its components, generalized across graph 
families, scaled beyond the regime where exact solving was 
practical, and remained interpretable through the recovered 
priority weights.

\section{Conclusion}

This work introduces MoSAIC, which is a learning-driven framework for linking optimization with heuristic scheduling in embedded systems. Scheduling is defined as an organized decision process, with CP-SAT optimal schedules used for supervision in learning reusable priority policies. Instead of using optimal schedules, their behavior is distilled into compact priority functions over structural properties such as rank, depth, and fanout, which are learned in an efficient manner through Bayesian Optimization. Following this definition, scheduling is defined as a prediction problem. A dataset is used to relate graph structures, motif patterns, and priority functions. This is used for generalization across various DAG structures. Large language models are used for prediction, and reinforcement learning is used for optimization. Experimental results have verified that MoSAIC achieves a smaller optimality gap than the classical heuristic algorithm HEFT, and meanwhile, the runtime overhead of MoSAIC is low. The learned policies are still interpretable, and ablation studies have verified the effectiveness of optimal supervision, heuristic recovery, and learning refinement. In summary, MoSAIC presents a promising direction for solving scheduling problems as a mapping from graph structure to priority-based policies, incorporating optimization, heuristics, and learning. In the future, we will extend MoSAIC to larger systems, dynamic environments, and more complex graph structures. This makes MoSAIC particularly suitable for embedded and resource-constrained systems where fast, adaptive, and interpretable scheduling decisions are required.


\begin{thebibliography}{00}

\bibitem{ref01}
M.~R. Garey and D.~S. Johnson,
\emph{Computers and Intractability: A Guide to the Theory of NP-Completeness}.
San Francisco, CA, USA: W.~H. Freeman, 1979.

\bibitem{ref02}
H.~Topcuoglu, S.~Hariri, and M.-Y. Wu,
``Performance-effective and low-complexity task scheduling for heterogeneous computing,''
\emph{IEEE Trans. Parallel Distrib. Syst.}, vol.~13, no.~3, pp.~260--274, Mar.~2002.

\bibitem{ref03}
Y.-K. Kwok and I.~Ahmad,
``Benchmarking the task graph scheduling algorithms,''
in \emph{Proc. Int. Parallel Process. Symp. (IPPS)}, 1999, pp.~531--537.

\bibitem{ref04}
L.~Perron and V.~Furnon,
``OR-Tools CP-SAT Solver,''
Google Research, 2020.

\bibitem{ref05}
H.~Mao, M.~Alizadeh, I.~Menache, and S.~Kandula,
``Resource management with deep reinforcement learning,''
in \emph{Proc. 15th ACM Workshop Hot Topics Netw. (HotNets)}, 2016, pp.~50--56.

\bibitem{ref06}
A.~Mirhoseini \emph{et al.},
``Device placement optimization with reinforcement learning,''
in \emph{Proc. 34th Int. Conf. Mach. Learn. (ICML)}, 2017, pp.~2430--2439.

\bibitem{ref07}
Y.~Bengio, A.~Lodi, and A.~Prouvost,
``Machine learning for combinatorial optimization: A methodological tour d'horizon,''
\emph{Eur. J. Oper. Res.}, vol.~290, no.~2, pp.~405--421, 2021.

\bibitem{ref08}
L.-N. Pouchet,
``PolyBench: The polyhedral benchmark suite,''
2012.

\bibitem{ref09}
B.~Reagen \emph{et al.},
``MachSuite: Benchmarks for accelerator design and customized architectures,''
in \emph{Proc. IEEE Int. Symp. Workload Characterization (IISWC)}, 2014, pp.~110--119.

\bibitem{ref10}
R.~Milo \emph{et al.},
``Network motifs: Simple building blocks of complex networks,''
\emph{Science}, vol.~298, no.~5594, pp.~824--827, Oct.~2002.

\bibitem{ref11}
M.~Milan,
``DAG-it-Right: A benchmark suite for DAG scheduling,''
GitHub repository. [Online]. Available: \texttt{https://github.com/MehradMilan/DAG-it-Right}

\bibitem{ref14}
T.~D.~Braun \emph{et al.},
``A comparison of eleven static heuristics for mapping a class of independent tasks onto heterogeneous distributed computing systems,''
\emph{J. Parallel Distrib. Comput.}, vol.~61, no.~6, pp.~810--837, 2001.

\bibitem{ref15}
O.~Sinnen,
\emph{Task Scheduling for Parallel Systems}.
Hoboken, NJ, USA: Wiley, 2007.

\bibitem{ref16}
P.~Laborie, J.~Rogerie, P.~Shaw, and P.~Vilím,
``IBM ILOG CP optimizer for scheduling,''
\emph{Constraints}, vol.~23, no.~2, pp.~210--250, 2018.

\bibitem{ref17}
J.~Hooker,
\emph{Integrated Methods for Optimization}.
New York, NY, USA: Springer, 2012.

\bibitem{ref18}
M.~Nazari, A.~Oroojlooy, L.~Snyder, and M.~Takáč,
``Reinforcement learning for solving the vehicle routing problem,''
in \emph{Proc. NeurIPS}, 2018.

\bibitem{ref19}
W.~Kool, H.~van Hoof, and M.~Welling,
``Attention, learn to solve routing problems!''
in \emph{Proc. ICLR}, 2019.

\bibitem{ref20}
B.~Shahriari, K.~Snoek, R.~P.~Adams, and H.~Larochelle,
``Taking the human out of the loop: A review of Bayesian optimization,''
\emph{Proc. IEEE}, vol.~104, no.~1, pp.~148--175, 2016.

\bibitem{ref21}
E.~J.~Hu, Y.~Shen, P.~Wallis, Z.~Allen-Zhu, Y.~Li, S.~Wang, L.~Wang, and W.~Chen,
``LoRA: Low-rank adaptation of large language models,''
in \emph{Proc. ICLR}, 2022.

\bibitem{ref22}
I.~Bello, H.~Pham, Q.~V.~Le, M.~Norouzi, and S.~Bengio,
``Neural combinatorial optimization with reinforcement learning,''
in \emph{Proc. ICLR Workshop}, 2017.

\bibitem{ref23}
T.~Kwok and I.~Ahmad,
``Dynamic critical-path scheduling: An effective technique for allocating
task graphs to multiprocessors,''
\emph{IEEE Trans. Parallel Distrib. Syst.}, vol.~7, no.~5,
pp.~506--521, 1996.

\bibitem{ref24}
P.~G. Paulin and J.~P. Knight,
``Force-directed scheduling for the behavioral synthesis of ASICs,''
\emph{IEEE Trans. Computer-Aided Design of Integrated Circuits and Systems},
vol.~8, no.~6, pp.~661--679, Jun.~1989.

\bibitem{ref25}
X.~Yan and J.~Han,
``gSpan: Graph-based substructure pattern mining,''
in \emph{Proc. IEEE Int. Conf. Data Mining (ICDM)}, 2002, pp.~721--724.

\end{thebibliography}
\end{document}